\documentclass{aa}
\begin{document}
\title{On the origin of the system PSR B\,1757$-$24/SNR G\,5.4$-$1.2}

\author{V.V.Gvaramadze\inst{1,2,3}\thanks{{\it Address for
correspondence}: Krasin str. 19, ap. 81, Moscow 123056, Russia
(vgvaram@sai.msu.ru)}}

\institute{Sternberg Astronomical Institute, Moscow State
University, Universitetskij Pr. 13, Moscow 119992, Russia \and
E.K.Kharadze Abastumani Astrophysical Observatory, Georgian
Academy of Sciences, A.Kazbegi ave. 2-a, Tbilisi 380060, Georgia
\and Abdus Salam International Centre for Theoretical Physics,
Strada Costiera 11, PO Box 586, 34100 Trieste, Italy}

\date{Received 2 June 2003/ Accepted 16 October 2003}

\titlerunning{On the origin of the system PSR B\,1757$-$24/SNR G\,5.4$-$1.2}
\authorrunning{Gvaramadze}

\abstract{A scenario for the origin of the system PSR
B\,1757$-$24/supernova remnant (SNR) G\,5.4$-$1.2 is proposed. It
is suggested that both objects are the remnants of a supernova
(SN) that exploded within a pre-existing bubble blown-up by a
runaway massive star (the SN progenitor) during the final
(Wolf-Rayet) phase of its evolution. This suggestion implies that
(a) the SN blast centre was significantly offset from the
geometric centre of the wind-blown bubble (i.e. from the centre of
the future SNR), (b) the bubble was surrounded by a massive
wind-driven shell, and (c) the SN blast wave was drastically
decelerated by the interaction with the shell. Therefore, one can
understand how the relatively young and low-velocity pulsar PSR
B\,1757$-$24 was able to escape from the associated SNR
G\,5.4$-$1.2 and why the inferred vector of pulsar transverse
velocity does not point away from the geometric centre of the SNR.
A possible origin of the radio source G\,5.27$-$0.9 (located
between PSR B\,1757$-$24 and the SNR G\,5.4$-$1.2) is proposed. It
is suggested that G\,5.27$-$0.9 is a lobe of a low Mach number
($\simeq 1.7$) jet of gas outflowing from the interior of
G\,5.4$-$1.2 through the hole bored in the SNR's shell by the
escaping pulsar. It is also suggested that the non-thermal
emission of the comet-shaped pulsar wind nebula originates in the
vicinity of the termination shock and in the cylindric region of
subsonically moving shocked pulsar wind. The role of magnetized
wind-driven shells (swept-up during the Wolf-Rayet phase from the
ambient interstellar medium with the regular magnetic field) in
formation of elongated axisymmetric SNRs is discussed. 
           }

\maketitle

%---------------------------------------------------------------------
\section{Introduction}
%---------------------------------------------------------------------
%
The pulsar PSR \object{B\,1757$-$24} is one of a growing number of
neutron stars associated with SNRs. Although this radio pulsar was
found  well outside the shell of the SNR \object{G\,5.4$-$1.2}
(Manchester et al. \cite{man}), until recently there was no doubt
about their physical association (Frail \& Kulkarni \cite{fra},
Manchester et al. \cite{manc}). The high-resolution VLA image
presented by Frail \& Kulkarni (\cite{fra}) shows a comet-shaped
flat-spectrum nebula stretched behind the pulsar towards
G\,5.4$-$1.2. This nebula was interpreted as a
ram-pressure-confined pulsar wind nebula, whose elongated
morphology is due to the high-velocity motion of the pulsar
through the interstellar medium.

Assuming that the pulsar was born in the apparent (geometric)
centre of G\,5.4$-$1.2 and that the true age of the pulsar,
$t_{\rm p}$, is equal to its characteristic spin-down age, $\tau =
P/2\dot{P} = 1.55\times 10^4$ yr, where $P=125$ ms and $\dot{P}
=1.28\times 10^{-13} \, {\rm s} \, {\rm s}^{-1}$ are,
respectively, the spin period and the period derivative, Frail \&
Kulkarni (\cite{fra}; see also Manchester et al. \cite{manc})
estimated the transverse velocity of the pulsar, $v_{\rm p} \geq
1900 \, d_{5} \, {\rm km}\,{\rm s}^{-1}$, where $d_{5}$ is the
distance to the system in units of 5 kpc (see Sect.\,2).

The above estimate could be somewhat reduced if the SN exploded in
a density-stratified medium: in this case, the actual SN explosion
site does not coincide with the geometric centre of the SNR, but
is shifted towards the high-density region (e.g. Gulliford
\cite{gul}). The offset is significant only if the characteristic
scale of the stratification is a small fraction of the radius of
the SNR. In this case the SNR's shell acquires a considerable
elongation. The elongation of the SNR G\,5.4$-$1.2, however, is
moderate. Frail et al. (\cite{frai}) used the Kompaneets
(\cite{com}) solution for a strong explosion in an
exponentially-stratified medium to fit the shape of G\,5.4$-$1.2
and found that the latter is consistent with the shape of a blast
wave expanding into the medium with the exponential scale height
less than one third of the diameter of the SNR. The model
presented by Frail et al. (\cite{frai}) allowed them to explain
the ``incorrect" orientation of the inferred line of pulsar proper
motion (the comet-shaped nebula produced by the moving pulsar does
not point back to the geometric centre of G\,5.4$-$1.2) and to
show that the SN blast centre could be offset towards the present
position of PSR B\,1757$-$24 by up to about a half of the radius
of the SNR. The pulsar transverse velocity implied by the new
possible location of the SN blast centre is $\simeq 1400 \, d_{5}
\, {\rm km}\,{\rm s}^{-1}$, that corresponds to the pulsar proper
motion of $\mu \simeq 60 \,{\rm mas} \, {\rm yr}^{-1}$ (Frail et
al. \cite{frai}).

Subsequent observations of the pulsar wind nebula separated by a
6.7 yr period did not show any appreciable changes in its
structure, putting a $5 \, \sigma$ upper limit on the pulsar
westward proper motion and the transverse velocity, respectively,
$\mu \leq 25 \, {\rm mas} \, {\rm yr}^{-1}$ and $v_{\rm p} \leq
590 \, d_5 \, {\rm km} \, {\rm s}^{-1}$ (Gaensler \& Frail
\cite{gae}). The discrepancy between the ``measured" transverse
velocity and the implied one is even greater if one adopts a $2 \,
\sigma$ upper limit on the westward pulsar motion derived by
Thorsett et al. (\cite{tho}) by combining their interferometric
proper motion measurements of PSR B\,1757$-$24 with the data taken
from Gaensler \& Frail (\cite{gae}), $\mu \leq 6.8 \, {\rm mas} \,
{\rm yr}^{-1}$, or
\begin{equation}
v_{\rm p} \leq v_{\rm p} ^{\rm max} = 160 \, d_{5} \,{\rm
km}\,{\rm s}^{-1} \, .
\label{eq1}
\end{equation}
Thorsett et al. (\cite{tho}) interpreted this discrepancy as an
indication of an equally large discrepancy between the kinematic
age of the system, $t_{\rm kin} =R_{\rm p} /v_{\rm p}$, where
$R_{\rm p}$ is the distance traveled by the pulsar from its
birth-place, and the characteristic age of the pulsar (cf.
Gaensler \& Frail \cite{gae}). The latter discrepancy constitutes
one of two arguments proposed by Thorsett et al. against the
physical association between PSR B\,1757$-$24 and the SNR
G\,5.4$-$1.2. The second argument against the association is the
``incorrect" orientation of the implied pulsar proper motion.
Thorsett et al. pointed out that a sharp density gradient across
the SNR [required by the model of G\,5.4$-$1.2 by Frail et al.
(\cite{frai})] ``without local inhomogeneities that disturb the
circular symmetry seems daunting", and concluded that the
(implied) proper motion direction is ``a serious problem for the
association hypothesis".

In this paper we propose a scenario for the origin of the system
PSR B\,1757$-$24/SNR G\,5.4$-$1.2 based on the idea that both
objects are the remnants of a cavity SN explosion of a moving
massive star. In this case, the offset of the SN blast centre from
the geometric centre of G\,5.4$-$1.2 could be maximum (i.e.
comparable with the radius of the SNR) even if the ambient
interstellar medium is homogeneous (Gvaramadze
\cite{gva2a},\cite{gva2b}; cf. Gvaramadze \& Vikhlinin
\cite{gvavik}, Gvaramadze \cite{gva3}). Our scenario implies a
much lower kinematic age of the system and naturally explains the
orientation of the comet-shaped pulsar wind nebula. In Sect.\,2 we
review the relevant observational data on the system PSR
B\,1757$-$24/SNR G\,5.4$-$1.2, while Sect.\,3 contains a scenario
for its origin. Sect.\,4 deals with some issues related to the
content of the paper. Sect.\,5 summarizes the work.

\section{System PSR B\,1757$-$24/SNR G\,5.4$-$1.2: observational data}

The 327 MHz image of the SNR G\,5.4$-$1.2 by Frail et al.
(\cite{frai}) shows a nearly circular region of diffuse emission
(of radius of $15{\farcm}5$), bounded from the west side by a
limb-brightened wing facing the Galactic plane. There are also
indications of a weaker and more amorphous east wing. Both wings
protrude in the north-south direction well beyond the area of
diffuse emission and remind the flanks of a barrel-like SNR (see
Sect.\,3.2 and Sect.\,4), so that the general structure of the SNR
is elongated rather than circular. The elongated (barrel-like)
structure of G\,5.4$-$1.2 could also be derived from the
polarization observations of this SNR: Milne et al. (\cite{mil})
found that the magnetic field is tangential around the wings of
the remnant and continues beyond the visible part of the shell,
nearly parallel to the Galactic plane.

Near to the west edge of G\,5.4$-$1.2 lies a bright, edge-darkened
compact source \object{G\,5.27$-$0.9}. This source is connected to
G\,5.4$-$1.2 by a bridge of emission. The VLA image of
G\,5.27$-$0.9 by Frail \& Kulkarni (\cite{fra}) resolves a small
protrusion at the west edge of this source into a comet-shaped
nebula, stretching for about $30^{\arcsec}$ east of PSR
B\,1757$-$24 until it merges with the radio source G\,5.27$-$0.9.
The elongated nebula does not point back to the geometric centre
of the SNR G\,5.4$-$1.2 but misses it by about $5^{\arcmin}$ to
the north (Frail et al. \cite{frai}).

The bright west wing of G\,5.4$-$1.2 shows a trend of a steepening
in the radio spectral index in either direction from the line
drawn though the pulsar and the bridge of emission connecting the
SNR with the radio source G\,5.27$-$0.9 (Frail et al.
\cite{frai}). The only available estimate of the spectral index
for G\,5.27$-$0.9, $\alpha= +0.2$ (Caswell et al. \cite{cas}),
suggests that the radio emission from this source is thermal.
However, the detection of linear polarization in G\,5.27$-$0.9
(Frail \& Kulkarni \cite{fra}) indicates that the emission is
rather non-thermal (see also Caswell et al. \cite{cas}). The radio
emission of the comet-shaped nebula is characterized by a flat
spectrum ($\alpha \simeq 0$) and a high degree of polarization
(Frail \& Kulkarni \cite{fra}), and therefore is also non-thermal.

Optical observations of a field containing G\,5.4$-$1.2 did not
reveal any features related to the radio shell of this SNR (Zealey
et al. \cite{zea}).

Recent {\it Chandra X-ray Observatory} observations of a
$4^{\arcmin} \times 4^{\arcmin}$ region of the west radio wing did
not detect the X-ray emission from this part of the SNR's shell,
but they led to the discovery of an X-ray counterpart to the
comet-shaped radio nebula associated with the pulsar (Kaspi et al.
\cite{kas}). The length of the X-ray tail is $\simeq
20^{\arcsec}$. It is likely that the tail X-ray emission is
non-thermal.

The estimates of the age of the system are very uncertain. The
maximum westward offset of the SN blast centre allowed by a model
of G\,5.4$-$1.2 and the upper limits on the pulsar westward motion
(see Sect.\,1) constrain the kinematic age of the system
\begin{equation}
t_{\rm kin} \geq R_{\rm p} /v_{\rm p} ^{\rm max} \, .
\label{eq2}
\end{equation}
Although, in principle, $v_{\rm p}$ could be as large as $590
\, {\rm km} \, {\rm s}^{-1}$, in the following we adopt for
$v_{\rm p}$ the maximum value allowed by the  ``worse" ($2 \,
\sigma$) upper limit on the pulsar westward motion, $v_{\rm p} =
160 \, {\rm km} \, {\rm s}^{-1}$, to show that even in this
``unfavourable" case the association between PSR B\,1757$-$24 and
G\,5.4$-$1.2 could be real (in Sect.\,4 we show, however,  that
the uncertainties in the pulsar transverse velocity do not affect
the main conclusions of the paper). The kinematic age should be
compared with the spin-down time-scale,
\begin{equation}
t_{\rm sd} = {2\over n-1} \left[1-\left({P_0 \over P}\right)^{n-1}
\right] \tau \, , \label{eq3}
\end{equation}
where $n$ is the braking index and $P_0$ is the initial spin
period. Note that all braking indices measured for young radio
pulsars are less than 3. It is plausible that the braking index of
PSR B\,1757$-$24 is also less than 3 and therefore $t_{\rm sd}
> \tau$ (provided that $P_0 << P$): the smaller $n$ the larger the
discrepancy between $t_{\rm sd}$ and $\tau$.

For $R_{\rm p} =23 \, d_{5}$ pc (Frail et al. \cite{frai}) and
assuming that $P_0 \simeq 0.5$ ms (i.e. the minimum spin period
allowed by the neutron star stability), one has from
(\ref{eq1}),(\ref{eq2}) and (\ref{eq3}) that $t_{\rm kin} (\geq
9\tau ) \simeq t_{\rm sd}$ if $n\leq 1.1$, i.e. for the braking
indices smaller than the smallest one ever measured for pulsars
(cf. Thorsett et al. \cite{tho}). On the other hand, in the case
of an off-centred cavity SN explosion $R_{\rm p}$ could be as
small as $\simeq 9 \, d_{5}$ pc (see Sect.\,3.1). In this case
$t_{\rm kin} (\geq 3.5\tau ) \simeq t_{\rm sd}$ if $n\leq 1.6$
(for $P_0 =0.5$ ms) or $n\leq 1.4$ (for $P_0 =5$ ms), i.e. for the
indices comparable with the braking index measured for the Vela
pulsar ($n=1.4\pm 0.2$; Lyne et al. \cite{lyn}).

Note also that $n<3$ is not the sole reason for the discrepancy
between $t_{\rm p}$ and $\tau$. The true age of a pulsar could be
larger than the characteristic spin-down age if the braking torque
acting on the pulsar grows with time (e.g. due to the secular
increase of the magnetic dipole moment; e.g. Blandford \& Romani
\cite{bla}) or is episodically enhanced (e.g. due to the
interaction between the pulsar's magnetosphere and the density
inhomogeneities in the ambient medium; e.g. Gvaramadze \cite{gva1}
and references therein).

In the following we assume that $t_{\rm p} =t_{\rm kin} =3.5\tau
\simeq 5.4\times 10^4$ yr.

The distance to the system PSR B\,1757$-$24/SNR G\,5.4$-$1.2 is
also poorly constrained. The kinematic method suggests that the
SNR is at a distance $>4.3$ kpc (Frail et al. \cite{frai}), while
the Cordes \& Lazio (\cite{cor}) model for the Galactic electron
density distribution gives the distance to the pulsar of $\simeq
5.1\pm 0.5$ kpc. Both estimates are not inconsistent with each
other. In what follows we adopt a distance to the system of $5$
kpc.

\section{System PSR B\,1757$-$24/SNR G\,5.4$-$1.2: a scenario for the origin}

We suggest that the origin of the system PSR B\,1757$-$24/SNR
G\,5.4$-$1.2 is connected to a SN explosion within a bubble
blown-up by the moving SN progenitor star during the Wolf-Rayet
(WR) phase of its evolution. We also suggest that the SN blast
centre was significantly offset towards the west edge of the
bubble due to the proper motion of the SN progenitor, and that the
SN blast wave was drastically decelerated by the interaction with
a pre-existing massive wind-driven shell. These suggestions imply
that PSR B\,1757$-$24 was born close to the current position of
the west edge of G\,5.4$-$1.2, and that the distance traveled by
the pulsar is much smaller than that allowed by the model of the
SNR by Frail et al. (\cite{frai}).

\subsection{Supernova progenitor and its processed ambient medium}

Let us explain why we believe that the pre-SN was a WR star (i.e.
the zero-age main-sequence (MS) mass of the SN progenitor was
$\geq 20 \, M_{\odot}$; e.g. Vanbeveren et al. \cite{van}) and
that the SN exploded within the WR bubble, but not in the bubble
created during the preceding MS phase. In our reasoning we proceed
from the fact that a young neutron star (born with a moderate kick
velocity of appropriate orientation) can overrun the shell of the
associated SNR only if: a) the SN exploded within a pre-existing
bubble surrounded by a massive (see Sect.\,3.2) shell; b) the SN
explosion site was significantly offset from the centre of the
bubble (e.g. Gvaramadze \cite{gva2a},\cite{gva2b}; cf. Shull et
al. \cite{shu}; Arzoumanian et al. \cite{arz}). However, it is
unlikely that these conditions can be fulfilled for the MS
bubbles. Indeed, simple estimates show that massive stars (unless
they are very massive and/or very slowly-moving) explode outside
their MS bubbles (Brighenti \& D'Ercole \cite{bri}). Moreover, the
MS bubbles usually stall and their shells disappear well before
the massive stars enter the subsequent evolutionary phases
(Brighenti \& D'Ercole \cite{bri}). On the other hand, if a
massive star ended its evolution as a WR star, the energetic WR
wind could create a new large-scale bubble, whose supersonic
expansion drives a shell of swept-up interstellar matter during
the relatively short WR phase. Besides, it is the short duration
of the WR phase that implies that even a runaway massive star
could explode within its WR bubble, while the stellar motion could
cause a considerable offset of the SN blast centre from the centre
of the bubble [see Fig.\,4 of Gruendl et al. (\cite{gru}) for a
good illustration of this effect; see also Arnal (\cite{arn})].

The proper motion of the SN progenitor star causes it to escape
from the bubble blown-up during the MS phase, so that the WR wind
interacts directly with the unperturbed interstellar medium. The
mass of the shell swept up by the end of the WR phase is $M_{\rm
sh}  = (4\pi / 3) R_{\rm sh} ^3 \rho _{\rm ISM}$, where $R_{\rm
sh}$ is the radius of the shell, $\rho _{\rm ISM} =1.3 m_{\rm H}
n_{\rm ISM}$, $n_{\rm ISM}$ is the number density of the ambient
interstellar medium and $m_{\rm H}$ is the mass of a hydrogen
atom. We stress that the (massive) wind-driven shell is a crucial
ingredient of our scenario since it is the interaction of the SN
blast wave with the shell that results in the abrupt deceleration
of the blast wave and that, in turn, allows the pulsar born with a
moderate kick velocity to overrun the SNR's shell. It is clear
that the larger the mass of the shell the stronger the
deceleration of the blast wave and the smaller the expansion
velocity of the resulting SNR. To estimate $M_{\rm sh}$ we need to
know $n_{\rm ISM}$ and $R_{\rm sh}$.

The number density could be crudely evaluated by comparing the
observed minimum size of the radio nebula ahead of the moving
pulsar with the theoretically predicted one,
\begin{equation}
R_{\rm n} \, = \, \kappa R_{0} \, = \, \kappa (|\dot{E}| /4\pi c
\rho _{\rm ISM} v_{\rm p} ^2 )^{1/2} \, ,
\label{eq4}
\end{equation}
where $R_{0}$ is the stand-off distance, $\kappa \simeq 1.26$ is a
parameter which shows that the radio emission ahead of the pulsar
comes from a layer of finite thickness of $\simeq (\kappa
-1)R_{0}$ (Bucciantini \cite{buc}), $|\dot{E}|$ is the spin-down
luminosity of the pulsar and $c$ is the speed of light; for the
sake of simplicity we assumed here that the pulsar is moving in
the plane of the sky and that the pulsar wind is isotropic. For
$R_{\rm n} = 3.6\times 10^{-2} d_{5}$ pc (Gaensler \& Frail
\cite{gae}) and $|\dot{E}| \simeq 2.6 \times 10^{36} \, {\rm erg}
\, {\rm s}^{-1}$, one has from Eq.\,(\ref{eq4}) that $n_{\rm ISM}
\simeq 1.0 \, {\rm cm}^{-3}$.

Then we assume that at the moment of SN explosion the radius of
the WR bubble/shell was $R_{\rm sh} =20$ pc. For this
value of $R_{\rm sh}$ and using the estimate of
$n_{\rm ISM}$, one has $M_{\rm sh} \simeq 10^3 \, M_{\odot}$.

We also assume that the SN exploded near the west edge of the WR
bubble (cf. Gvaramadze \& Vikhlinin \cite{gvavik}, Gvaramadze
\cite{gva3}), on the line drawn through the tail behind the
pulsar. In this case $R_{\rm p} \simeq 9 \, d_5$ pc, i.e the SN
blast centre is about $6^{\arcmin}$ east of the current position
of the pulsar (or about $3.5 \, d_5$ pc behind the west edge of
the SNR G\,5.4$-$1.2). Correspondingly, $t_{\rm kin} \simeq 5.4
\times 10^4 \, {\rm yr} \, (\simeq 3.5 \tau)$ (see Sect.\,2). One
can also estimate the peculiar velocity of the progenitor star,
$v_{\star} \simeq R_{\rm sh}/t_{\rm WR} \simeq 65 \, {\rm km} \,
{\rm s}^{-1}$, where $t_{\rm WR} \simeq 3\times 10^5$ yr is the
duration of the WR phase (see, e.g., Vanbeveren et al.
\cite{van}), i.e. the SN progenitor was a runaway star (cf.
Bandiera \cite{ban}). Note that the well-known runaway O-type star
$\zeta$ Pup has a similar peculiar velocity.

\subsection{SNR G\,5.4$-$1.2}

Soon after the SN explosion the blast wave starts to interact with
the closest (west) part of the shell, while in the opposite
direction it freely expands through the low-density interior of
the WR bubble until it collides with the shell. The further
evolution of the blast wave depends on the mass distribution over
the pre-existing shell. The SN blast wave becomes radiative if the
shell column density (or the mass of the shell) exceeds a critical
value.

There exist two main factors that affect the mass distribution
over the shell. The first one is the large-scale density gradient
in the ambient interstellar medium (usually oriented perpendicular
to the Galactic plane). The role of this factor is obvious: the
denser the ambient medium the larger the column density of the
swept-up shell. The second factor is the large-scale interstellar
magnetic field (at low Galactic latitudes it is nearly parallel to
the Galactic plane). It causes transverse motions in the shell:
the swept-up matter flows from the magnetic poles of the shell to
the equator and thereby increases (up to ten times) the column
density in this region of the shell (Ferri{\`e}re et al.
\cite{fer}). These factors naturally define the symmetry axes of
the wind-driven shell (respectively, perpendicular and parallel to
the Galactic plane) and later on those of the SNR (cf. Gvaramadze
\cite{gva9}, \cite{gva2b}). The inhomogeneous mass distribution
over the shell results in that the SN blast wave first enters the
radiative stage near the magnetic equator, that in turn results in
the origin of an asymmetric and/or bilateral brightness
distribution over the SNR's shell. Moreover, the reduced column
density at the magnetic poles of the wind-driven shell results in
the elongation of the SNR along the direction of the local
large-scale interstellar magnetic field (see also Sect.\,4).

The north-south elongation and the east-west brightness asymmetry
of the SNR G\,5.4$-$1.2 suggest that both aforementioned factors
have an effect on this SNR and that the column density of the
pre-existing wind-driven shell was maximum to the
west\footnote{Note, however, that the asymmetric brightness of
G\,5.4$-$1.2 could also be caused by the proximity of the SN blast
centre to the west edge of the pre-existing shell.}. The spectral
index variations along the west wing of G\,5.4$-$1.2 (see
Sect.\,2) also suggest that in this direction the SN blast wave
encountered a region of enhanced (column) density [cf. Thorsett et
al. (\cite{tho}) and see references therein]. Therefore we expect
that the expansion velocity of G\,5.4$-$1.2 is minimum to the
west.

To estimate the expansion velocity of the SNR one can use the
results of numerical simulation of cavity SN explosions by
Tenorio-Tagle et al. (\cite{ten}), which showed that the SN blast
wave merges with the pre-existing wind-driven shell, and the
reaccelerated shell (now the SNR's shell) evolves into a
momentum-conserving stage, if the mass of the shell is larger than
about $50 M_{\rm ej}$, where $M_{\rm ej}$ is the mass of the SN
ejecta. For any reasonable initial mass of the SN progenitor,
$M_{\rm ej} \simeq 3.5-10 \, M_{\odot}$ (see Table\,1 of
Vanbeveren et al. \cite{van}). Thus, $M_{\rm sh} > 50 M_{\rm ej}$
and the SNR G\,5.4$-$1.2 is in the radiative stage\footnote{This
conclusion does not contradict the non-detection of optical
emission from G\,5.4$-$1.2 (see Sect.\, 2) in view of the large
foreground obscuration towards this SNR (Caswell et al.
\cite{cas}).}. The numerical simulations by Tenorio-Tagle et al.
(\cite{ten}) also showed that the reaccelerated shell acquires a
kinetic energy $E_{\rm kin} = \beta E$, where $E=10^{51}$ erg is
the SN energy and $\beta \simeq 0.1$. Thus, the initial expansion
velocity of the SNR G\,5.4$-$1.2 is $ \simeq (2\beta E/M_{\rm sh}
)^{1/2} \simeq 100 \, {\rm km} \, {\rm s}^{-1}$. The westward
expansion of the SNR, however,  could be even slower due to the
inhomogeneous mass distribution over the shell, so that the pulsar
moving in the same direction with a velocity exceeding that of the
SNR's shell can easily overrun the SNR.

\subsection{G\,5.27$-$0.9}

We now discuss the origin of the radio source G\,5.27$-$0.9. We
suggest that G\,5.27$-$0.9 is a lobe of a low Mach number jet of
gas outflowing from the interior of G\,5.4$-$1.2 through the hole
bored in the SNR's shell by the escaping pulsar.

While the pulsar is moving through the shell of the SNR it creates
a channel filled with hot, low-density gas of the SNR's interior.
After the pulsar overruns the SNR the gas starts to outflow
through the hole in the shell and forms a supersonic jet. The gas
velocity at the origin of the jet is $v_{\rm j} = [2(c_{\rm j} ^2
- c_{\rm ISM} ^2 )/(\gamma -1)]^{1/2} \simeq \sqrt {3} \, c_{\rm
j}$, where $c_{\rm j} \, (>> c_{\rm ISM})$ is the sound speed of
the gas escaping from the SNR,  $c_{\rm ISM}$ is the sound speed
of the ambient interstellar medium and $\gamma =5/3$ is the
specific heat ratio. The structure and the dynamics of supersonic
jets propagating through the ambient medium are mainly determined
by two parameters (measured at the origin of the jet): the jet
Mach number, ${\cal{M}}_{\rm j} =v_{\rm j} /c_{\rm j}$, and the
jet to ambient medium density ratio, ${\rho}_{\rm j} /\rho _{\rm
ISM}$ (see Norman et al. \cite{nor}). It is clear that in our case
${\cal{M}}_{\rm j} \simeq 1.7$ and ${\rho}_{\rm j} /\rho _{\rm
ISM} << 1$.

Numerical simulations conducted by Norman et al. (\cite{nor})
showed that a low Mach number (${\cal{M}}_{\rm j} \sim 1.5$) and
low-density jet ends in a gradually inflating and slowly-moving
lobe. The morphological similarity of this lobe (see Fig.\,10a of
Norman et al. \cite{nor}) and the radio nebula G\,5.27$-$0.9 (see
Fig.\,1b of Frail \& Kulkarni \cite{fra}) allows us to consider
the existence of inner bright spots in G\,5.27$-$0.9 and the
edge-darkened appearance of this nebula (whose radio emission is
likely due to the synchrotron losses of relativistic electrons
accelerated at the internal shocks and those injected in the
nebula by the pulsar) as indications that the jet has already
reached its maximum spatial extent (see Norman et al. \cite{nor}).
Therefore the pulsar moving along the jet axis was able to overrun
the lobe and now it travels through the interstellar medium.

\subsection{PSR B\,1757$-$24 and its comet-shaped nebula}

Is is clear that the proper motion of a neutron star born in an
off-centred cavity SN explosion could be oriented arbitrarily with
respect to the geometric centre of the associated SNR (Gvaramadze
\cite{gva2a},\cite{gva2b}; see also Bock \& Gvaramadze
\cite{boc}). Therefore one should not comment on why the
comet-shaped nebula produced by PSR B\,1757$-$24 does not point
back to the geometric centre of the SNR G\,5.4$-$1.2. We discuss
some points related to the origin of this nebula.

The supersonic motion of PSR B\,1757$-$24 through the interstellar
medium results in the origin of an elongated structure, where the
pulsar wind is swept back by the ram pressure. The region occupied
by the wind is bounded by a contact discontinuity, which
asymptotically becomes cylindrical with a characteristic radius
$R\simeq 0.85\, {\cal{M}}_{\rm p} ^{3/4} (1-0.85\, {\cal{M}}_{\rm
p} ^{-1/2} )^{-1/4} \, R_0$, where ${\cal{M}}_{\rm p} = v_{\rm p}
/c_{\rm ISM}$ (Bucciantini \cite{buc}). For $v_{\rm p} =160\, {\rm
km} \, {\rm s}^{-1}$ and assuming that the temperature of the
ambient interstellar medium is $\simeq 8\,000 \, {\rm K}$, one has
$R \simeq 6\, R_0$ (or $\simeq 7^{\arcsec} \, d_5 ^{-1}$), i.e.
the value a few times larger than the half-width of the
comet-shaped nebula. This implies that we see only a central
(axial) part of a much wider region filled with the pulsar wind
and that most of the pulsar wind is unobservable.

We suggest that the non-thermal X-ray emission of the cometary
tail behind the pulsar is due to the synchrotron losses of the
relativistic pulsar wind shocked at the termination shock, which
extends in the tail up to a distance of $L \simeq 1.29 \,
{\cal{M}}_{\rm p} R_0$ (see Bucciantini \cite{buc} and Fig.\,1
therein), and where the wind particles acquire non-zero pitch
angles. Indirect support for this suggestion comes from the
comparison of $L \simeq 16\, R_0$ or $\simeq 19^{\arcsec} \, d_5
^{-1}$ with the observed length of the X-ray tail of $\simeq
20^{\arcsec}$.

We also suggest that the (non-thermal) radio emission of the
comet-shaped nebula originates in the vicinity of the termination
shock and in a much more extended narrow cylindrical region of
subsonically moving shocked pulsar wind (cf. Bucciantini
\cite{buc}). This suggestion implies that in the absence of the
radio source G\,5.27$-$0.9 the radio tail would be much longer
than its X-ray counterpart [perhaps as long as the tail of the
radio nebula ``Mouse" (\object{G\,359.23$-$0.82}; Yusef-Zadeh \&
Bally \cite{yus}) powered by the young pulsar \object{PSR
J\,1747$-$2958} (whose spin characteristics are almost the same as
those of PSR B\,1757$-$24; Camilo et al. \cite{cam})].

Finally, we note that the cometary morphology of the radio tail
(the tail is wide close to the pulsar, then narrows, and then
gradually widens again until it merges with G\,5.27$-$0.9) could
be interpreted as an indication that the magnetic field of the
pulsar wind is responsible for the shaping of the pulsar wind
nebula.

\section{Discussion}

We now show that the uncertainties in the pulsar velocity do not
affect the main conclusions of the paper.

First,  we consider the possibility that $v_{\rm p}$ could be
larger than $160 \, {\rm km} \, {\rm s}^{-1}$ (see Sect.\,1 and
Sect.\,2).  Note that the larger $v_{\rm p}$ the smaller $t_{\rm
kin}$ and the smaller the discrepancy between the latter and
$\tau$. For $v_{\rm p}  \, = \, 590 \, {\rm km} \, {\rm s}^{-1}$
and $R_{\rm p} \simeq 9$ pc (Sect.\,3), one has $t_{\rm kin}
\simeq 1.50\times 10^4$ yr $\simeq \tau$. On the other hand, the
larger $v_{\rm p}$ the smaller $n_{\rm ISM}$ implied by
Eq.\,(\ref{eq4}) and the smaller $M_{\rm sh}$. Therefore one can
estimate the maximum value of $v_{\rm p}$ consistent with our
suggestion that the SN blast wave becomes radiative after it
encountered the pre-existing shell. For the minimum mass of the
shell of $\simeq 175 \, M_{\odot}$ (see Sect.\,3.2), one has
$n_{\rm ISM} \simeq 0.18 \, {\rm cm}^{-3}$ and $v_{\rm p} \simeq
380 \, {\rm km} \,{\rm s}^{-1}$. The latter estimate implies
$t_{\rm kin} \simeq 2.2\times 10^4$ yr, which is consistent with
$\tau$ if $n\simeq 2.4$ (i.e. for the braking index comparable
with that of the Crab pulsar). Note, however, that for $v_{\rm p}
= 380 \, {\rm km} \,{\rm s}^{-1}$ the length of the termination
shock $L$ estimated in Sect.\,3.4 is about 2 times larger than the
observed length of the X-ray tail associated with PSR
B\,1757$-$54.

Second, one can consider the possibility that $v_{\rm p} < 160 \,
{\rm km} \, {\rm s}^{-1}$. In this case, the smaller $v_{\rm p}$
the larger $n_{\rm ISM}$ and the larger $M_{\rm sh}$. Let us
assume that $v_{\rm p} = 120 \, {\rm km} \, {\rm s}^{-1}$ (cf.
Thorsett et al. \cite{tho}). From (\ref{eq4}) one has that $n_{\rm
ISM} =1.8 \, {\rm cm}^{-3}$, that corresponds to $M_{\rm sh}
\simeq 1.5\times 10^3 \, M_{\odot} >> 50 \, M_{\rm ej}$, i.e. the
SNR is in the radiative stage. The only ``unpleasant" consequence
of the reduction of $v_{\rm p}$ is the increase of $t_{\rm kin}$.
However, for $R_{\rm p} = 9$ pc, one has that $t_{\rm kin} (\simeq
4.7 \tau ) \simeq t_{\rm sd}$ if $n<1.4$ (for $P_0 =0.5$ ms) or
$n<1.2$ (for $P_0 =5$ ms), i.e. for the braking indices still
comparable with that of the Vela pulsar (recall that the
assumption that $n<3$ is not the only way to reconcile $t_{\rm
kin}$ and $\tau$).

In conclusion we discuss an issue related to our suggestion that
PSR B\,1757-24 and the SNR G\,5.4$-$1.2 are the remnants of a SN
explosion within a bubble blown-up by the moving SN progenitor
star during the WR phase of its evolution. Namely we briefly
discuss the origin of elongated axisymmetric SNRs (e.g.
\object{G\,3.7$-$0.2}, \object{G\,16.2$-$2.7},
\object{G\,296.5$+$10.0}, \object{G\,332.4$-$0.4},
\object{G\,356.3$-$1.5}), constituting a subclass of the more
general class of bilateral SNRs (e.g. Kestenev \& Caswell
\cite{kes}).

Recently Gaensler (\cite{gaen}) demonstrated that the bilateral
SNRs show a generic tendency to be aligned with the local
large-scale Galactic magnetic field. This tendency implies that
the regular interstellar magnetic field is responsible not only
for the bilateral symmetry of these SNRs, but also for the
elongated shape of some of them. On the other hand, it is known
that the tension associated with the interstellar magnetic field
cannot directly affect the shape of a typical SN blast wave to
cause it to be elongated (e.g. Manchester \cite{manch}). To
explain the origin of elongated SNRs, Gaensler (\cite{gaen})
suggested that the SN blast waves in these SNRs take on the shape
of wind bubbles blown-up by the SN progenitor stars during the MS
phase and distorted by the surrounding (regular) magnetic field
(cf. Arnal \cite{arn}). A main concern with this suggestion is
that the majority of massive stars explode outside their MS
bubbles (see Sect\,3.1). In principle, one cannot exclude that the
progenitors of some elongated SNRs were very slowly-moving massive
stars. But even in this case the stellar motion results in a
significant offset of the SN blast centre from the centre of the
MS bubble (for $v_{\star} = 1 \, {\rm km} \, {\rm s}^{-1}$ and
$t_{\rm MS} \simeq 10^7$ yr, this offset is $\simeq 10$ pc).
Therefore, it is likely that at least two elongated bilateral SNRs
(G\,296.5$+$10.0 and G\,332.4$-$0.4) with centrally located
stellar remnants have a different origin.

We propose that the elongated axisymmetric SNRs are the diffuse
remnants of SNe that exploded within WR bubbles surrounded by
magnetized wind-driven shells. As we already mentioned in
Sect.\,3.2, the regular ambient magnetic field modifies the
structure of wind-driven shells in such a way that the density
distribution over the shell acquires an axial symmetry with the
minimum column density at the magnetic poles. Thus the origin of
elongated axisymmetric SNRs could be attributed to the interaction
of the SN blast wave with a pre-existing axisymmetric WR shell,
whose orientation with respect to the Galactic plane is determined
by the orientation of the local large-scale interstellar magnetic
field [see Fig. 1, 4 and 6 of Gruendl et al. (\cite{gru}) for
several examples of axisymmetric shells created by WR stars; note
that these shells are aligned nearly parallel to the Galactic
plane]. Our proposal implies that at least some progenitors of SNe
resulting in the origin of elongated axisymmetric SNRs are massive
stars that ended their evolution as WR stars. The detailed
analysis of this problem will be presented elsewhere.

\section{Summary}

We have proposed a scenario for the origin of the system PSR
B\,1757$-$24/supernova remnant G\,5.4$-$1.2 based on the
suggestion that both objects are the remnants of a supernova that
exploded within a pre-existing bubble blown-up by a runaway
massive star (the supernova progenitor) during the final
(Wolf-Rayet) phase of its evolution. Our suggestion implies that
(a) the supernova blast centre was significantly offset from the
geometric centre of the wind-blown bubble (i.e. from the centre of
the future supernova remnant), (b) the bubble was surrounded by a
massive wind-driven shell, and (c) the supernova blast wave was
drastically decelerated by interaction with the shell. Therefore
one can understand how the relatively young and low-velocity
pulsar PSR B\,1757$-$24 was able to escape from the associated
supernova remnant G\,5.4$-$1.2 and why the inferred vector of
pulsar transverse velocity does not point away from the geometric
centre of the remnant. A possible origin of the radio source
G\,5.27$-$0.9 (located between PSR B\,1757$-$24 and G\,5.4$-$1.2)
was proposed. It has been suggested that this nebula is a lobe of
a low Mach number ($\simeq 1.7$) jet of gas outflowing from the
interior of G\,5.4$-$1.2 through the hole bored in the shell of
this supernova remnant by the escaping pulsar. We have discussed
the origin of the comet-shaped pulsar wind nebula and suggested
that the non-thermal emission of this nebula originates in the
vicinity of the termination shock and in the cylindric region of
subsonically moving shocked pulsar wind. We also discussed the
origin of elongated axisymmetric supernova remnants and suggested
that they are the diffuse remnants of supernova explosions within
pre-existing Wolf-Rayet bubbles surrounded by axisymmetric shells,
whose axes of symmetry are aligned parallel to the local
large-scale interstellar magnetic field.

\begin{acknowledgements}
I am grateful to R.Bandiera for the interesting and useful
correspondence and to the anonymous referee for useful suggestions
and comments. This work was partially supported by the Deutscher
Akademischer Austausch Dienst (DAAD).

\end{acknowledgements}

\end{document}